\documentclass[aps,prd,reprint,superscriptaddress]{revtex4-1}

\usepackage{graphicx}
\usepackage{amssymb, amsmath,mathtools,amsthm}
\newtheoremstyle{dotless}{}{}{\itshape}{}{\bfseries}{}{ }{}
\theoremstyle{dotless}
\makeatletter
\def\@endtheorem{\endtrivlist}
\makeatother
\newtheorem*{proposition*}{}

\usepackage[colorlinks=false,urlcolor=black]{hyperref}

\newcommand{\be}{\begin{equation}}
\newcommand{\bea}{\begin{align}}
\newcommand{\eea}{\end{align}}
\newcommand{\beq}{\begin{equation}}
\newcommand{\ee}{\end{equation}}
\newcommand{\eeq}{\end{equation}}

\newcommand{\mnras}{{MNRAS}}		
		
\newcommand{\pasj}{{PASJ}}

\def\ip{${\cal I}^+$}

\begin{document}


\title{Black hole jet power from impedance matching}



\author{Robert F. Penna}
\email[]{rpenna@mit.edu}
\affiliation{Department of Physics and Kavli Institute for Astrophysics and Space Research,
Massachusetts Institute of Technology, Cambridge, Massachusetts 02139, USA}


\date{\today}

\begin{abstract}

Black hole jet power depends on the angular velocity of magnetic field lines, $\Omega_F$.  Force-free black hole magnetospheres typically have $\Omega_F/\Omega_H \approx 0.5$, where $\Omega_H$ is the angular velocity of the horizon.  We give a streamlined proof of this result using an extension of the classical black hole membrane paradigm.  The proof is based on an impedance-matching argument between membranes at the horizon and infinity.  Then we consider a general relativistic magnetohydrodynamic simulation of an accreting, spinning black hole and jet.  We find that the theory correctly describes the simulation in the jet region.  However, the field lines threading the horizon near the equator have much smaller $\Omega_F/\Omega_H$ because the force-free approximation breaks down in the accretion flow. 

\end{abstract}


\maketitle

\section{Introduction}
\label{sec:intro}

The Blandford-Znajek (BZ) model \cite{1977MNRAS.179..433B} describes how spinning black holes can power jets.  The black hole's rotational energy is extracted by the interaction of an external magnetic field with the hole.  The BZ model's predictions depend on the angular velocity of magnetic field lines, $\Omega_F$.  One way to compute $\Omega_F$ is to build models for black hole magnetospheres.  The simplest realistic black hole magnetospheres are described by force-free electrodynamics (FFE), already a nonlinear set of differential equations.  
There exist analytic perturbative solutions for slowly rotating black holes
 \cite{1977MNRAS.179..433B,
 2008PhRvD..78b4004T,
 2013arXiv1303.1644B,
 2014arXiv1406.4936P,
 2014MNRAS.445.2500G,
 2015PhRvD..91f4067P,
Yang:2015ata,
 2015arXiv150404864P,
 2015arXiv150402113G} 
 and numerical solutions for rapidly rotating black holes 
 \cite{2001MNRAS.326L..41K,
  Tchekhovskoy:2009ba,
  Palenzuela:2010xn,
 2013ApJ...765..113C,
 2014ApJ...788..186N}.  
General relativistic magnetohydrodynamics (GRMHD) simulations describe the interaction of jets with accretion flows 
 \cite{2004MNRAS.350.1431K,
 2005MNRAS.359..801K,
 McKinney:2006tf,
 2011MNRAS.418L..79T,
 McKinney:2012vh,
 2013MNRAS.436.3741P}.

In all cases (analytical and numerical, FFE and GRMHD), one finds that realistic jets have $\Omega_F/\Omega_H\approx 0.4-0.5$, where $\Omega_H$ is the angular velocity of the horizon.  This universality begs an explanation.  It is also striking that $\Omega_F/\Omega_H=0.5$ happens to be the value that maximizes jet power in the BZ model.   The black hole membrane paradigm suggests an explanation for both observations \cite{phinney1983,1986bhmp.book.....T,1992MNRAS.254..192O,2006PASJ...58.1047O,2009PASJ...61..971O,2012PASJ...64...50O,2015PASJ..tmp..217O}.  In this picture, currents flowing in the magnetosphere are closed off by currents flowing on the horizon and the sphere at infinity.  The ratio $\Omega_F/\Omega_H$ is the circuit efficiency.  The horizon and infinity have the same surface resistivity, so they achieve near perfect impedance matching for most field geometries.  This explains why the circuit efficiency is $\Omega_F/\Omega_H\approx 0.5$.

Recently, we have given an action principle formulation of the membrane at infinity \cite{Penna:2015qta}, extending an earlier action principle formulation of the membrane at the horizon \cite{Parikh:1997ma}.  This motivates us to revisit the impedance-matching argument.  We give special attention to the electromagnetic boundary condition at infinity.  Then we consider a GRMHD simulation of an accreting black hole with spin $a/M=0.9$.  The simulation spontaneously develops a jet with $\Omega_F/\Omega_H\approx 0.4$.  We check that FFE is a good approximation in the jet region.  We show that the boundary conditions defining the membrane at infinity are approximately satisfied inside the jet, at radii beyond the stagnation surface, which we define below.  We conclude that the membrane impedance-matching argument correctly explains the simulation data.

The description of the horizon and future null infinity, \ip, as resistive membranes is valid for observers who remain in the black hole exterior.  We will not consider observers who fall in the black hole.  However, the membrane paradigm can be extended to this case.  Each observer defines a causal diamond of events with which they are in causal contact.  Each observer sees the future boundary of their causal diamond as a membrane with similar properties to the membranes we use in this paper.

Our paper is organized as follows.  In Sec. \ref{sec:result} we give a streamlined version of the impedance-matching argument, in Sec. \ref{sec:bc} we discuss the electromagnetic boundary condition at infinity, in Sec. \ref{sec:grmhd} we apply the theory to our GRMHD simulation, and in Sec. \ref{sec:conc} we summarize and discuss future directions.

\section{Impedance matching}
\label{sec:result}

The Kerr metric in Boyer-Lindquist (BL) coordinates is
\begin{align}
ds^2 &= -(1-2Mr/\rho^2)dt^2 - (4Mar \sin^2\theta/\rho^2)dtd\phi\notag\\
&+(\rho^2/\Delta)dr^2+\rho^2 d\theta^2\notag\\
&+(r^2+a^2+2Ma^2r\sin^2\theta/\rho^2)\sin^2\theta d\phi^2,
\end{align}
where 
\begin{align}
\Delta &= r^2-2Mr+a^2,\\
\rho^2 &= r^2+a^2\cos^2\theta,
\end{align}
and $M$ and $a$ are the mass and spin of the black hole.  The event horizon is at $r_H=M^2+\sqrt{M^2-a^2}$ and the angular velocity of the horizon is $\Omega_H=a/(2Mr_H)$. The zero-angular momentum observer (ZAMO) frame is \cite{1972ApJ...178..347B}
\begin{align}
e^{\hat{t}} &= \alpha dt,\\
e^{\hat{r}} &= \rho/\Delta^{1/2}dr,\\
e^{\hat{\theta}} &= \rho d\theta,\\
e^{\hat{\phi}} &= -\omega \Sigma/\rho dt
+\varpi d\phi,
\end{align}
where
\begin{align}
\alpha &= \frac{\rho}{\Sigma}\Delta^{1/2},\\
\omega &= \frac{2Mar}{\Sigma^2},\\
\Sigma^2&=(r^2+a^2)^2-a^2\Delta\sin^2\theta,\\
\varpi &= \frac{\Sigma}{\rho}\sin\theta.
\end{align}

An observer in the black hole exterior cannot receive signals from beyond the horizon or \ip.  These surfaces behave as membranes at the edge of space  \cite{1986bhmp.book.....T,1992MNRAS.254..192O,Parikh:1997ma,Penna:2015qta}.  They carry surface 3-currents, $\vec{j}_H$ and $\vec{j}_\infty$, which terminate the electromagnetic field in accordance with Gauss's law and Ampere's law.  The membrane at infinity has infinite area but finite charge (assuming standard boundary conditions on the electromagnetic field).  To avoid infinities, one computes the charge on ``stretched infinity,'' a surface at large but finite radius, and then takes the limit as stretched infinity approaches true infinity \cite{Penna:2015qta}.  A stretched horizon slightly outside the true horizon plays a similar role in regulating infinities at the horizon.  We assume throughout that infinite quantities have been regularized using the stretched horizon and stretched infinity in the standard way.  We use the ZAMO frame because it is the fiducial example of a family of observers who remain in the black hole exterior.

\begin{figure}
\includegraphics[width=\columnwidth]{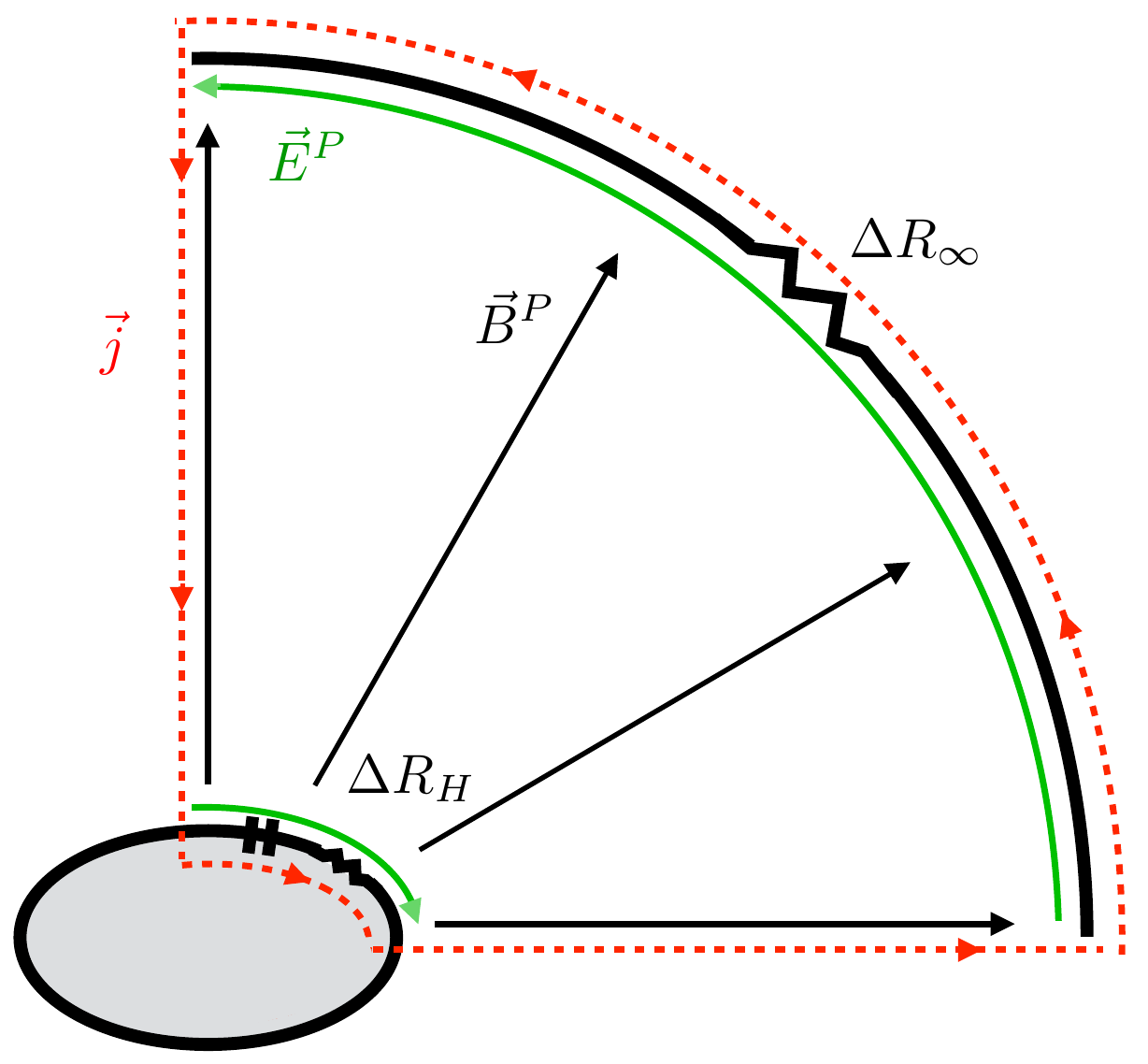}
\caption{Force-free black hole magnetosphere as a circuit.  Heavy black lines indicate the membrane at the stretched horizon and the membrane at stretched infinity.  In the magnetosphere, the current (dashed red) follows the poloidal magnetic field, $\vec{B}^P$.  On the membranes, the current follows the poloidal electric field, $\vec{E}^P$ (membrane Ohm's law).  The total current forms a closed circuit.  The black hole acts as a battery driving the circuit.  The field line angular velocity, $\Omega_F/\Omega_H$, is the circuit efficiency. It is fixed by the ratio between the resistive drop at infinity, $\Delta R_\infty$, and the resistive drop at the horizon, $\Delta R_H$.  This ratio is fixed by the angular distributions of the field lines at infinity and at the horizon [see Eq. \ref{eq:matching}].  As long as the field line distribution is fairly uniform, one has near perfect impedance matching and  $\Omega_F/\Omega_H\approx 0.5$.}
\label{fig:circuit}
\end{figure}

The 3-current on the membrane at the horizon is
\beq\label{eq:jH}
j_H^a = F^{a\hat{r}},
\eeq
and the 3-current on the membrane at infinity is
\beq\label{eq:jinfty}
j_\infty^a = -F^{a\hat{r}},
\eeq
where $\pm F^{a\hat{r}}=\pm F^{ab}e^{\hat{r}}_b$ is the electromagnetic field in the black hole exterior evaluated at the horizon and infinity. The membrane currents close off the currents in the magnetosphere and form closed circuits.  The problem of computing $\Omega_F/\Omega_H$ becomes a circuit problem (see Figure \ref{fig:circuit}).

The membranes are resistors, with surface resistivities  (see Sec. \ref{sec:bc}):
\beq\label{eq:377}
R_H=R_\infty = 377 \Omega.
\eeq
The black hole's rotation acts as a voltage source.  So we have a three-element circuit consisting of a battery and two resistors connected in series.

Let  $\vec{E}$ by the ZAMO frame electric field and let $\vec{E}_\parallel$ be its components parallel to the membranes.  The membrane electric fields, $\vec{E}_H$ and $\vec{E}_\infty$, are 2-vectors obtained by evaluating $\vec{E}_\parallel$ at the membranes.  They obey Ohm's laws,
\beq\label{eq:ohm}
\vec{E}_H = R_H \vec{j}_H, \quad
\vec{E}_\infty = R_\infty \vec{j}_\infty,
\eeq
with resistivities given by \eqref{eq:377}.

Assume a stationary, axisymmetric, force-free magnetosphere.  
The electric field is
\beq\label{eq:efield}
\vec{E} = -\vec{v}_F \times \vec{B}^P,
\eeq
where $\vec{B}^P$ is the poloidal magnetic field in the ZAMO frame, and 
\begin{align}
\vec{v}_F &= \frac{1}{\alpha}(\Omega_F-\omega)\varpi e_{\hat{\phi}},\label{eq:vF}\\
\Omega_F &= -F_{t\theta}/F_{\phi\theta},\label{eq:omegaF}
\end{align}
are the linear and angular velocities of magnetic field lines.

\begin{figure}
\includegraphics[width=\columnwidth]{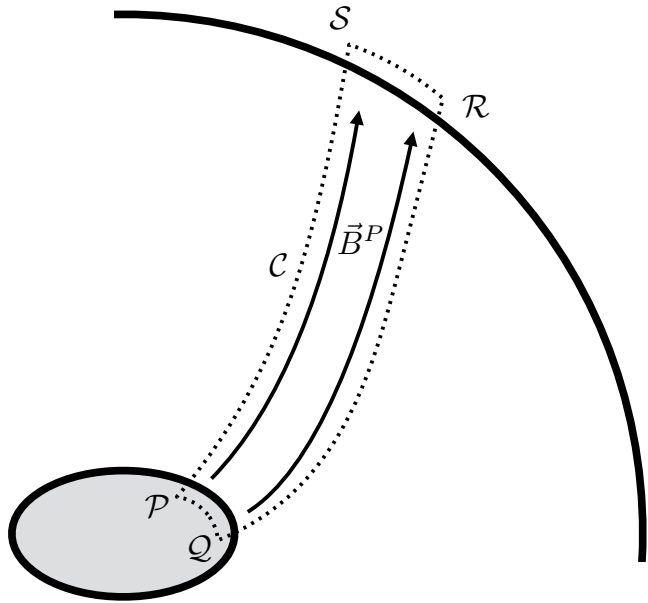}
\caption{The poloidal magnetic field, $\vec{B}^P$, extends from the stretched horizon to stretched infinity.  The closed curve, $\mathcal{C}$, runs along the field lines and closes off along the membranes.   The resistive drop across a membrane is proportional to the angular distance traversed by $\mathcal{C}$ across the membrane.  In this example, $\mathcal{RS}$ has a smaller angular size than $\mathcal{PQ}$, so these field lines have $\Delta R_\infty / \Delta R_H<1$ and $\Omega_F/\Omega_H<0.5$.  }
\label{fig:circuit1}
\end{figure}

Consider the electromotive force (EMF) generated by the black hole battery.  Let $\mathcal{C}$ be the closed curve in Fig. \ref{fig:circuit1}.  The curve begins at $\mathcal{Q}$ on the stretched horizon, rises along a magnetic field line to the point $\mathcal{R}$ on stretched infinity, runs along stretched infinity to $\mathcal{S}$, follows a different magnetic field line back to the stretched horizon at $\mathcal{P}$, and returns along the stretched horizon to the starting point. 
The total EMF around $\mathcal{C}$ is
\beq
\Delta V=\oint_{\mathcal{C}}\alpha \vec{E}\cdot {d\vec{l}}.
\eeq
Assume $\mathcal{C}$ is at rest with respect to BL coordinates, so its velocity with respect to ZAMOs is $\vec{v}=\vec{\beta}/\alpha$, where $\vec{\beta}=(0,0,-\omega)$.  $\mathcal{C}$ is a closed contour, so we can use Faraday's law to obtain  \cite{1986bhmp.book.....T} 
\beq
\Delta V=-\oint_{\mathcal{C}} \vec{\beta} \times \vec{B}\cdot d\vec{l}.
\eeq
The only contribution to this integral is from the segment of $\mathcal{C}$ along the horizon, so
\beq
\Delta V=-\int_{\mathcal{P}}^{\mathcal{Q}} \vec{\beta} \times \vec{B} \cdot d\vec{l}
=\frac{1}{2\pi}\Omega_H \Delta A_\phi,
\eeq
where $\Delta A_\phi$ is the magnetic flux threading the horizon between $\mathcal{P}$ and $\mathcal{Q}$.  The hole's rotation acts as a battery supplying the EMF, $\Delta V$.  This drives the current around the circuit.

Now consider the resistors at the horizon and infinity.   (There is no resistance in the magnetosphere itself because $\vec{E}^P$ is perpendicular to $\vec{B}^P$ for force-free magnetospheres.)  The EMF is balanced by voltage drops, 
\begin{align}\label{eq:VH}
\Delta V_H &= \int_{\mathcal{P}}^{\mathcal{Q}} \alpha \vec{E}\cdot d\vec{l},\\
\Delta V_\infty &= \int_{\mathcal{R}}^{\mathcal{S}} \alpha \vec{E}\cdot d\vec{l},\label{eq:Vinf}
\end{align}
across the horizon and infinity.  The total voltage drop is $\Delta V=\Delta V_H + \Delta V_\infty$.  Ohm's law \eqref{eq:ohm} gives
\begin{align}
\Delta V_H &= I\Delta R_H,\label{eq:resisH}\\
\Delta V_\infty & = I\Delta R_\infty,\label{eq:resisinf}
\end{align}
where $I$ is the integrated current,
\begin{align}
\Delta R_H &= R_H \left[\frac{\rho \Delta \theta}{2\pi \varpi}\right]_H,\label{eq:dRH}\\
\Delta R_\infty &= R_\infty \left[\frac{\rho \Delta \theta}{2\pi \varpi}\right]_\infty,\label{eq:dRinf}
\end{align}
and $\Delta \theta_H$ and $\Delta \theta_\infty$ are the angular lengths of $\mathcal{PQ}$ and $\mathcal{RS}$.   So the voltage drop on each membrane is proportional to $377 \Omega$ and the angular distance  $\mathcal{C}$ traverses across the membrane.

We can relate the voltage drops to $\Omega_F/\Omega_H$ because the electric field is induced by the rotating magnetic field.  Plugging \eqref{eq:efield} into \eqref{eq:VH} and \eqref{eq:Vinf} gives
\begin{align}
\Delta V_H &=\frac{1}{2\pi}(\Omega_H - \Omega_F)\Delta A_\phi,\label{eq:dVH}\\
\Delta V_\infty &=\frac{1}{2\pi} \Omega_F\Delta A_\phi.\label{eq:dVinf}
\end{align}
Dividing \eqref{eq:dVinf} by \eqref{eq:dVH} gives
\beq
\frac{\Omega_F}{\Omega_H-\Omega_F}
= \frac{\Delta V_{\infty}}{\Delta V_H}
= \frac{\Delta R_{\infty}}{\Delta R_H},
\eeq
where in the last step we used \eqref{eq:resisH}-\eqref{eq:resisinf}.  Now solving for $\Omega_F/\Omega_H$ gives
\beq\label{eq:eff}
\Omega_F/\Omega_H = \frac{\Delta R_\infty /\Delta R_H}{1+\Delta R_\infty /\Delta R_H}.
\eeq
This shows that $\Omega_F/\Omega_H$ is precisely the circuit efficiency.  Just as for ordinary circuits, maximum power output is achieved at the load when the efficiency is $50\%$.  When the efficiency is low, most of the power generated by the black hole battery is dissipated in the horizon.  When the efficiency is high, the load power is small because the overall resistance of the circuit is large.   Maximum jet power corresponds to perfect impedance matching between the horizon and infinity.

Dividing \eqref{eq:dRH} by \eqref{eq:dRinf} gives 
\beq\label{eq:matching}
\Delta R_\infty / \Delta R_H 
= \frac{[\Delta \theta\rho/\varpi]_\infty}{[\Delta \theta\rho/\varpi]_H}
= \frac{[A_{\phi,\theta}\varpi/\rho ]_H}{[ A_{\phi,\theta}\varpi/\rho]_\infty},
\eeq
where in the second equality we have used the fact that $A_\phi$ is constant on field lines.  The factors of $377\Omega$ have dropped out.  All that remains is a dependence on the angular distribution of $A_{\phi}$ at the horizon and infinity.  If the field is fairly uniform at the horizon and infinity, then $\Omega_F/\Omega_H \approx 0.5$.    FFE magnetospheres tend to maximize jet power (up to order unity factors) because they tend to relax to roughly uniform field distributions at the horizon and infinity.

\begin{figure}
\includegraphics[width=\columnwidth]{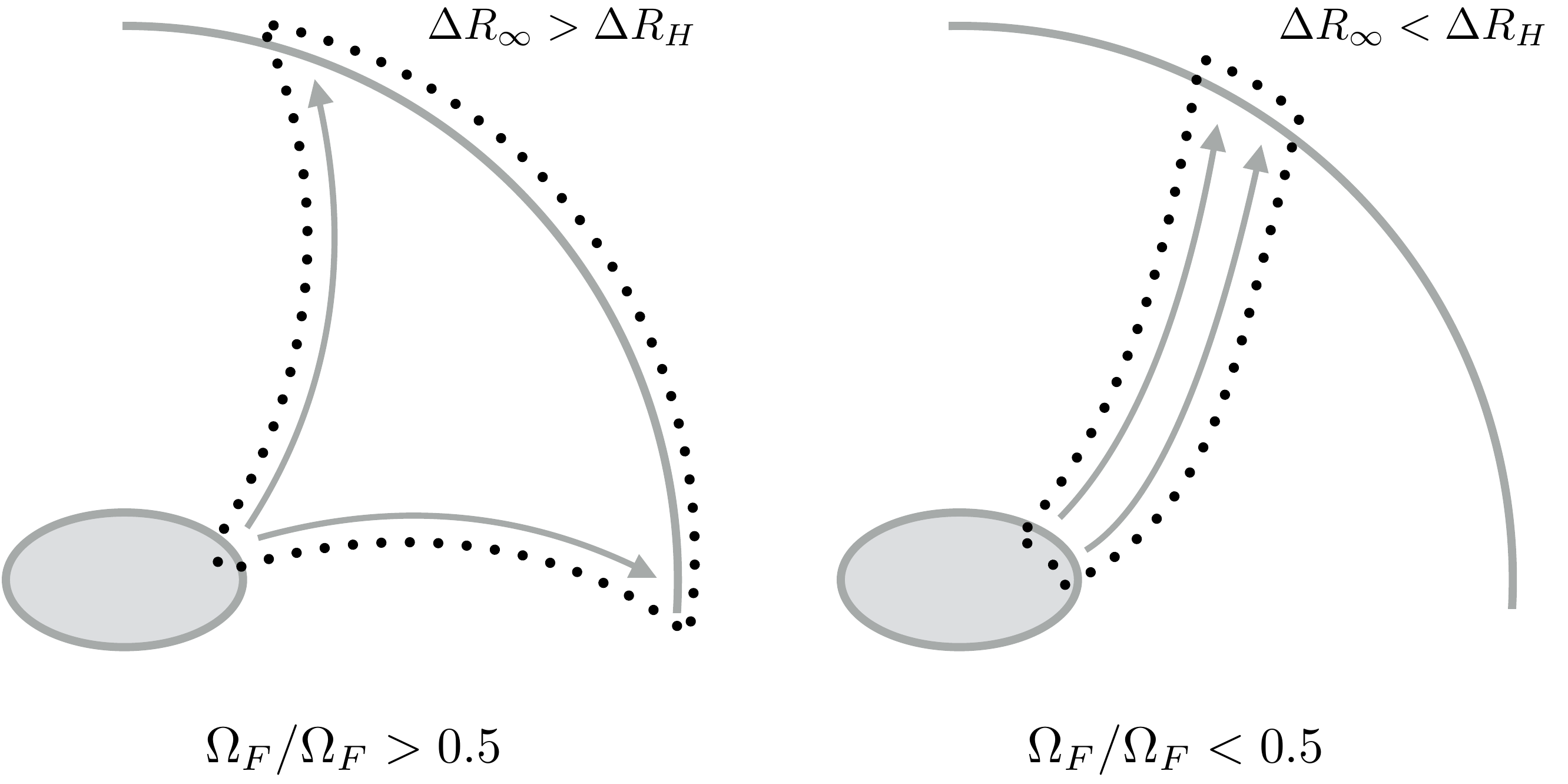}
\caption{The magnetosphere on the left is overloaded: there is a larger resistive drop across infinity than across the horizon, so  $\Omega_F/\Omega_H>0.5$.  The magnetosphere on the right is underloaded, so $\Omega_F/\Omega_H < 0.5$.}
\label{fig:cases}
\end{figure}

Equations \eqref{eq:eff}--\eqref{eq:matching} imply a simple rule for deciding whether a force-free magnetosphere will have $\Omega_F/\Omega_H$ greater than or less than $0.5$.  If the magnetic field lines have diverging angular separation as they approach infinity, then the voltage drop is larger at infinity than at the horizon and $\Omega_F/\Omega_H > 0.5$.  If the field lines have converging angular separation as they approach infinity, then the voltage drop is smaller at infinity than at the horizon and $\Omega_F/\Omega_H<0.5$. Both cases are illustrated in Fig. \ref{fig:cases}.

\subsection{Examples}
\label{sec:examples}

The simplest FFE magnetosphere is the slowly rotating split monopole \cite{1977MNRAS.179..433B}.  The field lines have constant angular separation, $\Delta \theta_H = \Delta \theta_\infty$.  So Eqs. \eqref{eq:eff}-\eqref{eq:matching} give $\Omega_F/\Omega_H=0.5$, as expected.  

There are no analytical FFE solutions for split monopoles at high spins, but \eqref{eq:eff}--\eqref{eq:matching} imply that split monopoles must have $\Omega_F/\Omega_H\approx 0.5$ at all spins.   This is consistent with numerical solutions for split-monopole fields at high spins \cite{2001MNRAS.326L..41K,Tchekhovskoy:2009ba,2010PhRvD..82d4045P,2013ApJ...765..113C,2014ApJ...788..186N} and  perturbative analytical split-monopole solutions extrapolated to high spins \cite{2008PhRvD..78b4004T,2015PhRvD..91f4067P}.

The next simplest solution is the slowly rotating paraboloidal field \cite{1977MNRAS.179..433B}.
The distribution of flux at infinity is
\beq
A_\phi = r(1-\cos\theta)/2,
\eeq
and the distribution at the horizon is 
\beq
A_\phi = 2\log 2-(1+\cos\theta)\log(1+\cos\theta).
\eeq
Plugging into \eqref{eq:eff}-\eqref{eq:matching} gives $\Omega_F$.  We find perfect agreement with a direct computation of $\Omega_F=-F_{t\theta}/F_{\phi\theta}$ at the horizon:
\begin{align}
\Omega_F/\Omega_H &= [\sin^2\theta(1+\log(1+\cos\theta))]\notag\\
&\times[4\log2+\sin^2\theta+(\sin^2\theta\notag\\
&-2(1+\cos\theta))\log(1+\cos\theta)]^{-1}.
\end{align}
This is a nontrivial check that \eqref{eq:eff}-\eqref{eq:matching} are correct.   

There are no analytical FFE solutions for paraboloidal fields at high spins, but  \eqref{eq:eff}--\eqref{eq:matching} imply that they must have $\Omega_F/\Omega_H$ close to (but slightly less than) $0.5$ at all spins. 

\section{Boundary conditions at infinity}
\label{sec:bc}

The crux of the argument in Sec. \ref{sec:result} is the claim that \ip\ has a dual description as a resistive membrane with $R_\infty = 377 \Omega$.  This claim has a two-line proof.  We assume the electromagnetic field satisfies the outgoing boundary condition 
\beq\label{eq:bczamo}
\vec{E}_\parallel = -\hat{n} \times \vec{B}_\parallel
\eeq
at infinity.  Then we use the definition of the membrane current, $\vec{j}_\infty =  -\hat{n} \times \vec{B}_\parallel$, to obtain
\beq
\vec{j}_\infty = \vec{E}_\parallel,
\eeq
which is Ohm's law on the membrane. Restoring units gives the surface resistivity $R_\infty = 1/(c\epsilon_0) = 377\Omega$. 
So $R_\infty = 377 \Omega$ is essentially equivalent to the boundary condition \eqref{eq:bczamo}.  We expect the impedance-matching argument to apply to all force-free fields which satisfy \eqref{eq:bczamo}.  

For stationary, axisymmetric FFE, the outgoing boundary condition at infinity \eqref{eq:bczamo} is equivalent to
\beq\label{eq:bcznajek}
B_T = -\Omega_F A_{\phi,\theta} \sin\theta,
\eeq
where $B_T=\sqrt{-g} F^{r\theta}$ is the toroidal magnetic field and $A_\phi$ is the magnetic flux function.   Equation \eqref{eq:bcznajek} is completely analogous to the Znajek condition \cite{1977MNRAS.179..457Z} at the horizon (which may be understood as an ingoing boundary condition).

To see the equivalence of \eqref{eq:bczamo} and \eqref{eq:bcznajek}, note that at infinity
\begin{align}
B_T & = r\sin\theta F_{\hat{r}\hat{\theta}} = r\sin\theta B^{\hat{\phi}},\\
\Omega_F &=  \frac{v_F^{\hat{\phi}}}{r\sin\theta},\\
A_{\phi,\theta} &= r^2\sin\theta F_{\hat{\theta}\hat{\phi}} = r^2\sin\theta B^{\hat{r}}.
\end{align}
Plugging into \eqref{eq:bcznajek} gives $B^{\hat{\phi}} = -v_F^{\hat{\phi}} B^{\hat{r}}$.  Using $\vec{E} = -\vec{v}_F\times \vec{B}$ gives $B^{\hat{\phi}}=E^{\hat{\theta}}$.  This is one component of \eqref{eq:bczamo}.  The other component is trivial because $E^{\hat{\phi}}=0$ in stationary, axisymmetric FFE.  So \eqref{eq:bczamo} and \eqref{eq:bcznajek} are equivalent boundary conditions at infinity in stationary, axisymmetric FFE.

Instead of a boundary condition at \ip, one can assume fall-off conditions for the electromagnetic field.  A standard choice is 
\beq
A_t= O(1/r), \quad A_r = 0, \quad A_\theta = O(1), \quad A_\phi = O(1).
\eeq
This is motivated by the fact that electromagnetic waves near \ip\ have the general form $A_\theta = C e^{ik(t-r)}$.  This implies that waves near \ip\ satisfy
\beq
E^{\hat{\theta}} = -F_{t\theta}/r = -ik A_\theta/r = F_{r\theta}/r = B^{\hat{\phi}}.
\eeq
A similar calculation gives $E^{\hat{\phi}} = - B^{\hat{\theta}}$.  So boundary conditions \eqref{eq:bczamo} and \eqref{eq:bcznajek} are consistent with the usual fall-off conditions.

\subsection{Examples}

Consider again the split-monopole solution of  \cite{1977MNRAS.179..433B}. This solution has 
\begin{align}
A_\phi &= -C \cos\theta,\\
A_{\phi,\theta} &= C \sin\theta, \\
\Omega_F &= \frac{1}{2}\Omega_H, \\
B_T &= - \frac{1}{2}\Omega_H C\sin^2\theta.
\end{align}
The outgoing boundary condition at infinity \eqref{eq:bcznajek} is clearly satisfied. 

A more interesting case is the paraboloidal solution of \cite{1977MNRAS.179..433B}.  For $r/M\gg 1$, this solution has
\begin{align}
A_\phi &= \frac{C}{2}r(1-\cos\theta) + 2 CM(1-\log2),\\
A_{\phi,\theta} &= \frac{C}{2} r\sin\theta,\\
B_T &= -C\Omega_F r (1-\cos\theta).
\end{align}
 For general $\theta$, the outgoing boundary condition \eqref{eq:bcznajek} is not satisfied.  However, all of the field lines threading the event horizon reach \ip\ at $\theta=0$.  Expanding near $\theta=0$ gives
\begin{align}
A_{\phi,\theta} &= \frac{C}{2} r\theta+O(\theta^3),\\
B_T &= -\frac{C}{2}\Omega_F r \theta^2+O(\theta^4),
\end{align}
so the outgoing boundary condition \eqref{eq:bcznajek} is satisfied near $\theta=0$.  This is why the impedance-matching argument applies to the field lines threading the horizon.  The portion of the paraboloidal solution not threading the horizon should probably be regarded as unrealistic, a stand-in for the accretion disk or whatever other physics is supporting the jet.

\subsection{Finite-length jets}

The membrane at infinity is an idealization.  Real astrophysical jets have finite length.  The key property of the membrane at infinity is $R_\infty= 377 \Omega$.  We have shown that this is equivalent to the outgoing boundary condition \eqref{eq:bczamo}.    So for the purposes of the impedance-matching argument, we can move the membrane to any cutoff surface where \eqref{eq:bczamo} is approximately satisfied and apply the argument.  The key question is how far from the event horizon we need to go before  \eqref{eq:bczamo} becomes a good approximation.

At the horizon, the electromagnetic field satisfies the ingoing boundary condition
\beq\label{eq:bczamoH}
\vec{E}_\parallel = \hat{n} \times \vec{B}_\parallel.
\eeq
This is the same as \eqref{eq:bczamo} except for a sign flip.  The upshot is that $\vec{E}^P$ points in opposite directions at the horizon and infinity.  At some intermediate radius, $r_{\rm es}$, it vanishes.  This is the electromagnetic stagnation (ES) surface.  At the ES surface, the field changes its character from ingoing to outgoing.  We expect the outgoing boundary condition \eqref{eq:bczamo} to be a good approximation for $r\gg r_{\rm es}$.

\begin{figure}
\includegraphics[width=\columnwidth]{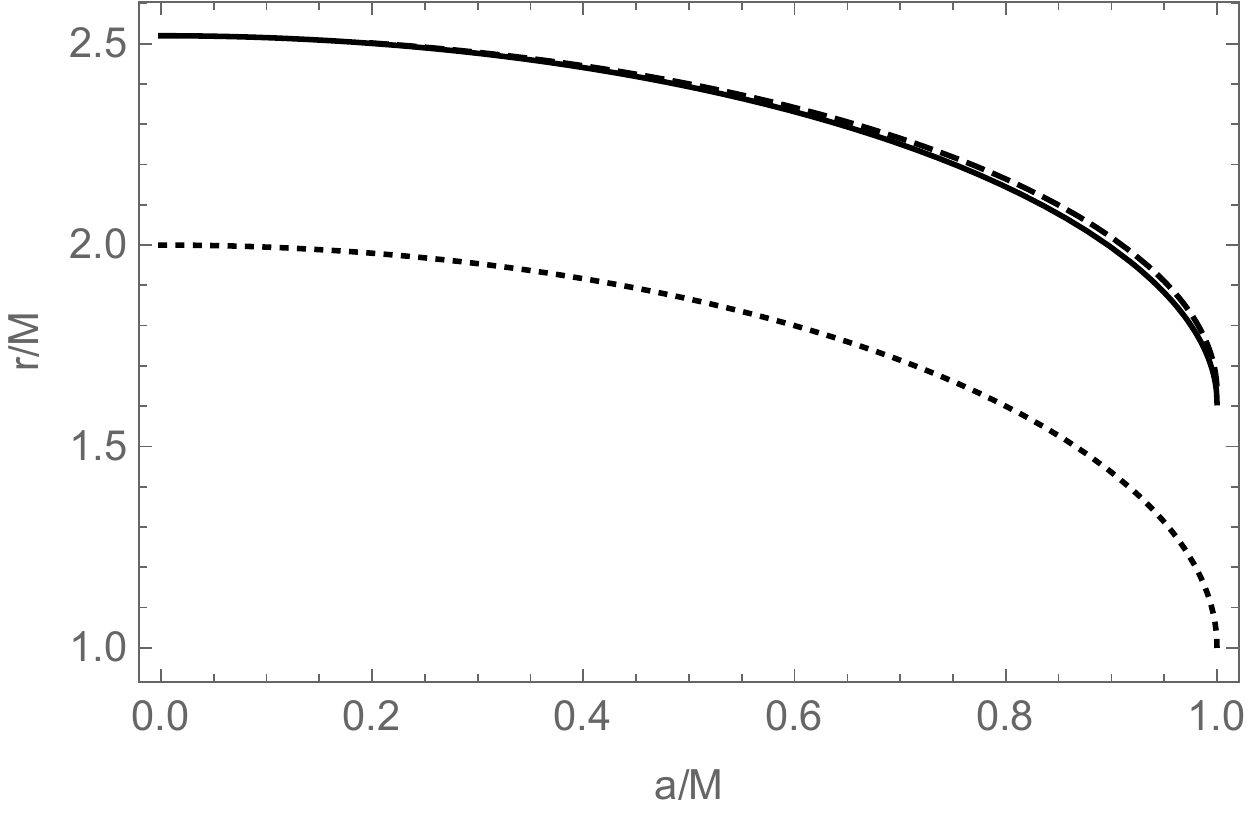}
\caption{Spin dependence of $r_{\rm es}$ at $\theta=0$ (solid) and $\theta=\pi/2$ (long-dashed) for $\Omega_F/\Omega_H=0.5$.  The short-dashed curve is $r_H$.}
\label{fig:ess}
\end{figure}

Equations \eqref{eq:efield}--\eqref{eq:vF} imply that the electric field vanishes in the ZAMO frame when
\beq
\omega(r_{\rm es},\theta) = \Omega_F.
\eeq
 Figure \ref{fig:ess} shows $r_{\rm es}$ as a function of $a/M$ for $\Omega_F/\Omega_H=0.5$.  For $a/M=0$,
\beq
r_{es}/M= 2^{4/3} \approx 2.5,
\eeq
a constant independent of $\theta$.  For $a/M=1$, the radius of the ES surface is $r_{es}/M\approx 1.6$, with a few percent variation between the poles and the equator.

So we expect the outgoing boundary condition \eqref{eq:bczamo} to be a good approximation even for $r/M < 100$.  The membrane at infinity can be moved inside the jet and the impedance-matching argument can be used to infer $\Omega_F/\Omega_H\approx 0.5$ for finite length jets.

The minimum distance from the black hole at which we can introduce a cutoff surface and define a ``membrane at infinity'' is the electromagnetic stagnation surface.  For astrophysical jets, there is also a maximum distance at which we can place the membrane, because the field far from the black hole is causally disconnected from the field near the horizon.  In particular, we should keep the membrane within the subsonic region of the flow.  \cite{2011CQGra..28m4007P} found numerical solutions for force-free BZ jets in which they fixed different boundary conditions on the electromagnetic field at $r/M \geq 16$.  They found that the jet power was insensitive to the choice of boundary condition at these radii.  One possible explanation is that these radii may already be in the supersonic region of the flow and, thus, causally disconnected from the field near the horizon.

\section{GRMHD simulation}
\label{sec:grmhd}

Accreting black holes are not force free.  In this section, we apply our results to a GRMHD simulation of an accreting black hole with $a/M=0.9$.  The simulation describes a 3+1-dimensional, turbulent accretion flow in the Kerr metric.  The accretion flow brings magnetic flux onto the hole and a BZ-like jet develops spontaneously.   We discuss $t-$ and $\phi-$averaged results from the steady state period of the simulation.   For a detailed description of the simulation see \cite{ 2013MNRAS.436.3741P}.

\begin{figure}
\includegraphics[width=\columnwidth]{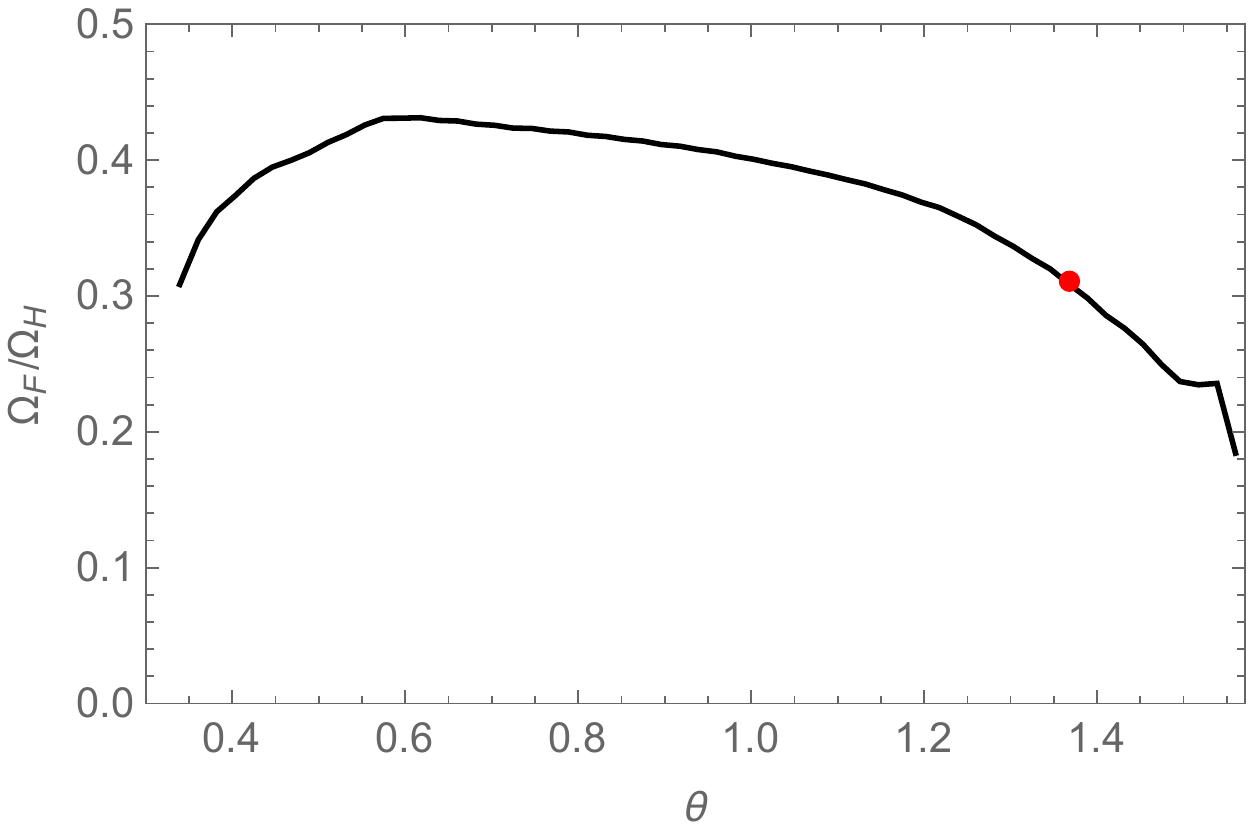}
\caption{$\Omega_F/\Omega_H$ at the horizon of an $a/M=0.9$ GRMHD simulation.  The red dot indicates the boundary between the jet (where the net energy flux is outward) and the accretion flow.}
\label{fig:omegaFa9}
\end{figure}

Define the jet to be the region of the simulation where the net energy flux is outwards.  
Figure \ref{fig:omegaFa9} shows $\Omega_F/\Omega_H$ at the horizon.  
In the jet, the angular velocity is $\Omega_F/\Omega_H \sim 0.4$, as expected from the impedance-matching argument.  Not all field lines threading the horizon are in the jet.  Field lines threading the horizon near the equator are in the accretion flow.  These field lines have smaller $\Omega_F/\Omega_H$.

\begin{figure}
\includegraphics[width=\columnwidth]{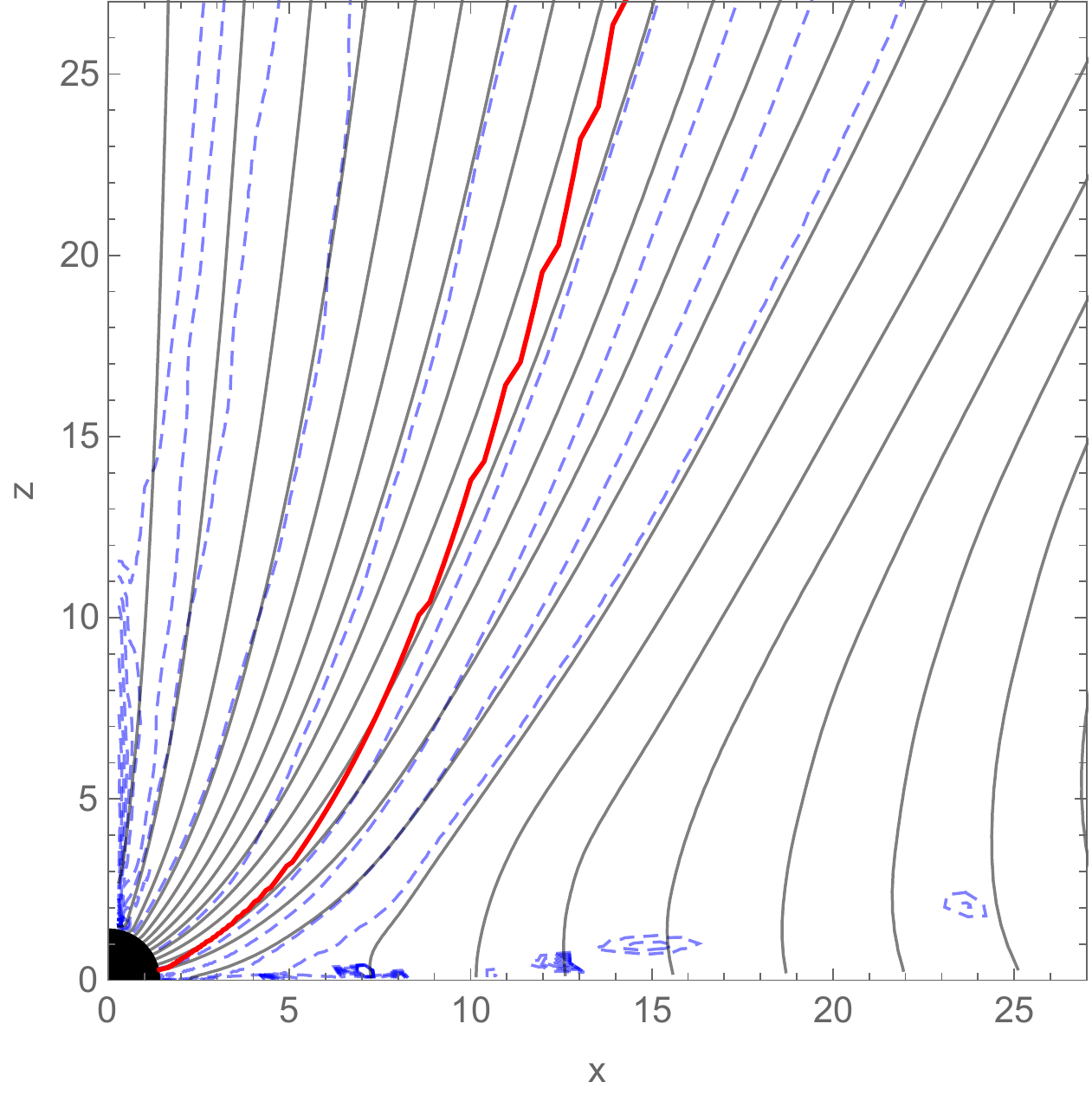}
\caption{Lines of constant $\Omega_F$ (dashed blue) overlaid on magnetic field lines (solid black).  The red curve is the boundary between the jet and the accretion flow.  The coordinates are $x=r/M \sin\theta$ and $z=r/M\cos\theta$.}
\label{fig:omegaF2da9}
\end{figure}

Figure \ref{fig:omegaF2da9} shows contours of $\Omega_F$ overlaid on magnetic field lines.  $\Omega_F$ is nearly constant on field lines.  It would be precisely constant on field lines in stationary, axisymmetric ideal MHD \cite{1978PhRvD..18.1809B}.  However, the simulations are not ideal MHD. They have (numerical) resistivity, as indeed they must: without resistivity, it would be impossible for gas to diffuse across field lines and there could be no accretion flow.  In addition, even the $t-$ and $\phi-$averaged simulation data is not completely stationary and axisymmetric.  Despite these caveats, $\Omega_F$ is roughly constant on field lines.

We would like to understand $\Omega_F/\Omega_H$ using the impedance-matching argument.  This argument relied on two key assumptions: (i) current flows along magnetic field lines, and (ii) the electromagnetic field satisfies the outgoing boundary condition \eqref{eq:bcznajek} at infinity.  

\begin{figure}
\includegraphics[width=\columnwidth]{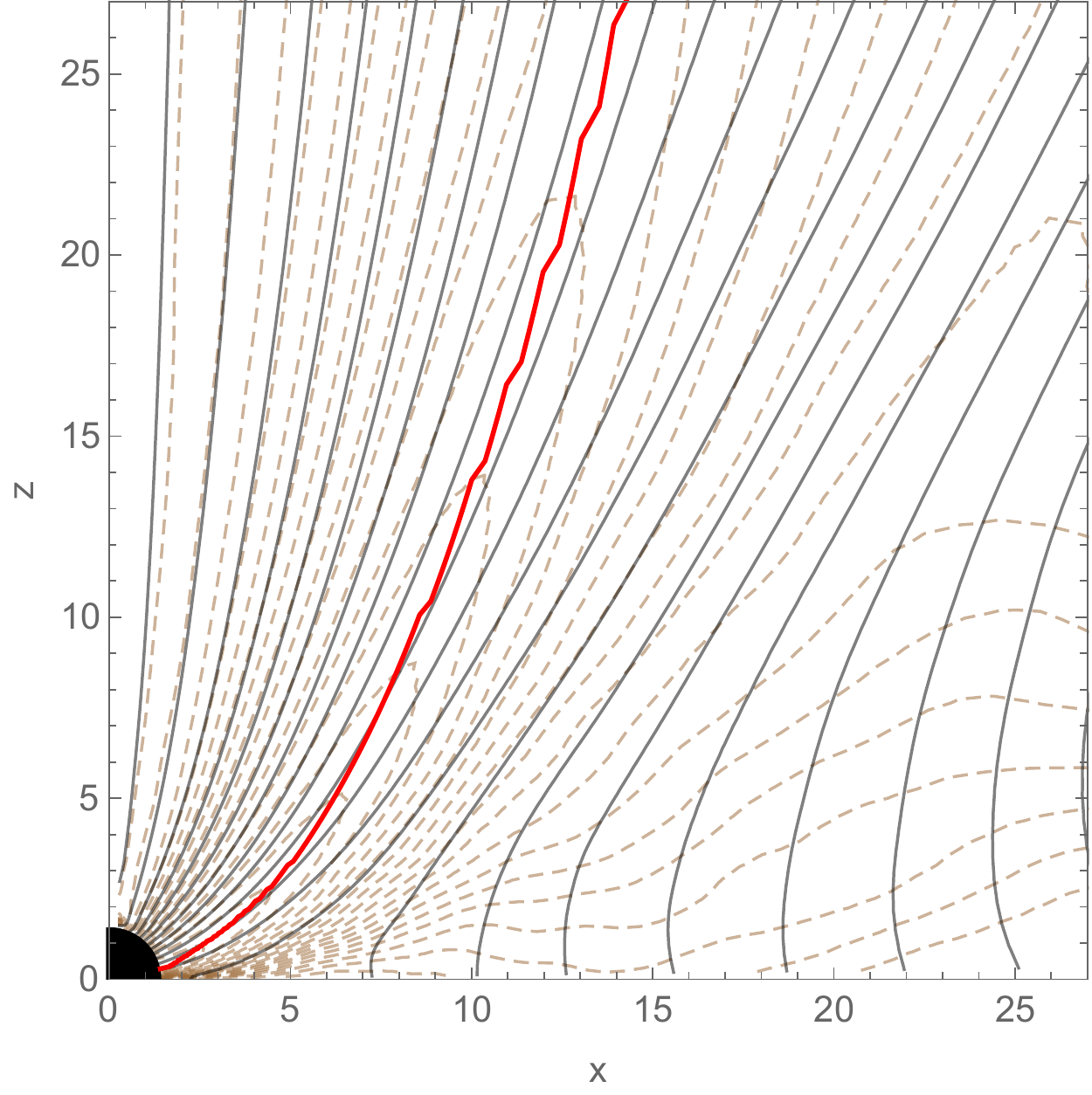}
\caption{Lines of constant $I$ (dashed brown) overlaid on magnetic field lines (solid black). The red curve is the boundary between the jet and the accretion flow.}
\label{fig:I2da9}
\end{figure}

First, consider assumption (i).  $I$ is constant on field lines in FFE but not in non-force-free MHD.   Figure \ref{fig:I2da9} shows contours of the GRMHD simulation's $I$ overlaid on magnetic field lines.  FFE [and, hence, assumption (i)] is a good approximation in the jet but not in the accretion flow.  This is as expected because the inertia of the gas is negligible in the jet but not the accretion flow.

\begin{figure}
\includegraphics[width=\columnwidth]{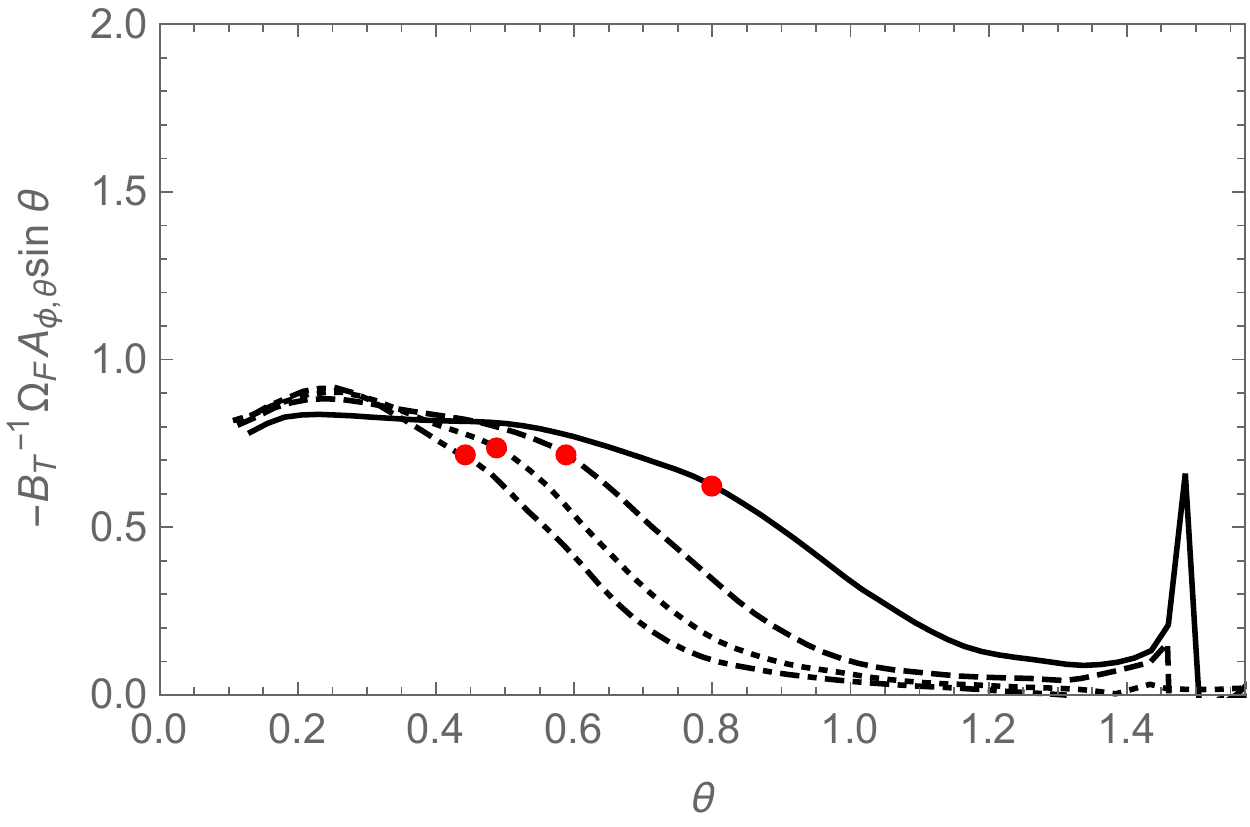}
\caption{The diagnostic plotted on the vertical axis is $1$ when the outgoing boundary condition \eqref{eq:bcznajek} is satisfied.  Curves correspond to $r/M=10$ (solid), $20$ (long-dashed), $30$ (short-dashed), and $40$ (dot-dashed).  Red dots indicate the boundary between the jet and the accretion flow.   The outgoing boundary condition is approximately satisfied in the jet but not in the accretion flow.}
\label{fig:ratioa9}
\end{figure}

Assumption (ii) is that the the electromagnetic field satisfies the outgoing boundary condition \eqref{eq:bcznajek} for $r/M\gg r_{\rm es}$, where $r_{\rm es}/M\approx 2$ is the radius of the electromagnetic stagnation surface.   Figure \ref{fig:ratioa9} shows the ratio
\beq
-\frac{\Omega_FA_{\phi,\theta}\sin\theta}{B_T}
\eeq
as a function of $\theta$ at several different radii.   This ratio is unity when the boundary condition \eqref{eq:bcznajek} is satisfied.  We find that \eqref{eq:bcznajek} is approximately satisfied in the jet, but not in the accretion flow.  This shows that the membrane ``at infinity'' can be moved to horizon-scale radii and the impedance-matching argument can be applied to the jet.  A different theory is needed to explain $\Omega_F/\Omega_H$ in the accretion flow.

\section{Discussion}
\label{sec:conc}

Our goal in this paper was to understand the physics controlling $\Omega_F/\Omega_F$ in black hole jets.  We gave a streamlined proof that all stationary, axisymmetric force-free jets have $\Omega_F/\Omega_H\approx 0.5$ (as long as the magnetic field distribution is not too pathological).  The proof led to a simple geometrical rule for determining whether $\Omega_F/\Omega_H$ is greater than or less than $0.5$: field lines with diverging (converging) angular separation tend to have  $\Omega_F/\Omega_H$ greater (less) than $0.5$.  We used the impedance-matching argument to interpret a GRMHD jet simulation.

Early models of black hole jets as circuits assumed $\Omega_F/\Omega_H$ is controlled by impedance matching between the horizon and an astrophysical load where the force-free approximation breaks down. In these models, $\Omega_F/\Omega_H$ depends on the ill-understood physics of the astrophysical load and the theory leaves ``to astrophysical model builders the horrendous task of trying to compute it \cite{1986bhmp.book.....T}.''  We identified the load with the membrane at infinity.  This simplifies the theory and it explains the universality of $\Omega_F/\Omega_H\approx 0.5$.  This universality traces back to the fact that the membranes at the horizon and infinity have the same resistivity: $R_H=R_\infty=377\Omega$.  

The membrane paradigm has been criticized as acausal \cite{2008ASSL..355.....P,2009JKPS...54.2503K}.  As a result, it has fallen into disfavor among astrophysical model builders (exceptions include \cite{1992MNRAS.254..192O,2006PASJ...58.1047O,2009PASJ...61..971O,2012PASJ...64...50O, 2013MNRAS.436.3741P,2015PASJ..tmp..217O}).  The impedance-matching argument applies to stationary solutions, so causality does not enter.  A more challenging problem is to understand how out-of-equilibrium states relax to stationary solutions.  The membrane paradigm description is acausal because the event horizon is teleological (its position depends on the entire future history of spacetime).

An interesting open problem  is to understand jets from boosted black holes and binary black holes (see \cite{Penna:2015qta} and references therein).  There are no analytical FFE solutions in these cases.  The impedance-matching argument described here may be useful.  (A version of this argument was applied to the binary problem in \cite{2011ApJ...742...90M}.)  There has been progress simulating boosted black holes in uniform vertical magnetic fields \cite{2011PNAS..10812641N}.  Uniform vertical fields do not satisfy the outgoing boundary condition \eqref{eq:bczamo}.  However, one could split the field into a uniform background field plus a perturbation caused by the motion of the black hole.  If the perturbation satisfies the outgoing boundary condition \eqref{eq:bczamo}, then it might be possible to adapt the impedance-matching argument to this component of the field \footnote{I thank Carlos Palenzuela and an anonymous referee for this suggestion}.

Solving FFE analytically requires a boundary condition at infinity.  The nature of this boundary condition has been a source of confusion \cite{ 2008PhRvD..78b4004T}.   In place of a boundary condition at infinity, \cite{2014arXiv1406.4936P, 2015PhRvD..91f4067P, 2015arXiv150404864P} use a ``convergence condition'' on the electromagnetic field to find FFE solutions.  Recently, this convergence condition was shown to be equivalent to a simple constraint on the fields at infinity \cite{2015arXiv150404864P}.  This constraint is a special case of the outgoing boundary condition \eqref{eq:bcznajek}.  In the future, it may be advantageous to use \eqref{eq:bcznajek} directly as the boundary condition at infinity when solving FFE (in the same way that the Znajek condition is used at the horizon).

It would be interesting to extend this analysis from black holes to neutron stars.  Neutron star jets can also be understood as circuits.  The role of the membrane at infinity is unchanged.  The membrane at the horizon is replaced by the neutron star surface.  One difference is that while the black hole membrane has a surface resistivity of $377\Omega$, a neutron star is a perfect conductor.  So to apply the circuit equations to a neutron star, we should take the limit $\Delta R_H\rightarrow 0$.  In this limit, Eq. \eqref{eq:eff} gives $\Omega_F = \Omega_H$, which means the field lines are forced to rotate at the same angular velocity as the neutron star surface, as expected.  Extending the circuit analogy further may help explain the jets produced by neutron stars as they undergo gravitational collapse to black holes \cite{2012PhRvD..86j4035L}.

\begin{acknowledgments}
I thank Vasily Beskin, Sam Gralla, Ramesh Narayan, Carlos Palenzuela, and Maria Rodriguez for comments.  This work was supported by a Pappalardo Fellowship in Physics at MIT.
\end{acknowledgments}

\bibliography{ms}

\end{document}